\documentclass[a4paper]{IEEEtran}
\usepackage{cite}
\usepackage{graphicx}
\usepackage{epsfig}
\usepackage{epstopdf}
\usepackage{algorithm}
\usepackage{algorithmic}
\usepackage{epstopdf}
\usepackage{epsfig}
\usepackage{textcomp}
\usepackage{url}
\usepackage{color}
\usepackage{pbox}
\usepackage{multirow}
\usepackage[style=base, font=footnotesize, labelfont={sf,bf}, margin=0cm]{caption}

\definecolor{darkgreen}{RGB}{255, 127, 0}
\newcommand{\modify}[1]{{#1}}

\newcommand{\remove}[1]{{}}
\ifCLASSINFOpdf
\else
\fi

\begin{document}

\title{Architectural Challenges and Solutions for Collocated LWIP - A Network Layer Perspective\vspace{-0.3cm}}
\author{\IEEEauthorblockN{Thomas Valerrian Pasca S, Amogh PC, Debashisha Mishra, Nagamani Dheeravath, Anil Kumar Rangisetti, \\Bheemarjuna Reddy Tamma and Antony Franklin A\\ }
\IEEEauthorblockA{Department of Computer Science and Engineering, Indian Institute of Technology Hyderabad, India\\
Email:[cs13p1002, cs15mtech01002, cs15mtech01003, cs11b011, cs12p1001,  tbr, antony.franklin]@iith.ac.in}\vspace{-0.9cm}}

\IEEEoverridecommandlockouts
\IEEEpubid{\makebox[\columnwidth]{Complete Version of the draft is available in proceeding of NCC 2017. }
    \hspace{\columnsep}\makebox[\columnwidth]{ }} 
\maketitle

\begin{abstract}
Achieving a tighter level of aggregation between LTE and Wi-Fi networks at the radio access network (a.k.a. LTE-Wi-Fi Aggregation or LWA) has become one of the most prominent solutions in the era of 5G to boost network capacity and improve end user\textquotesingle s quality of experience. LWA offers flexible resource scheduling decisions for steering user traffic via LTE and Wi-Fi links. In this work, we propose a Collocated LTE/WLAN Radio Level Integration architecture at IP layer (C-LWIP), an enhancement over 3GPP non-collocated LWIP architecture.  We have evaluated C-LWIP performance in various link aggregation strategies (LASs). A C-LWIP node (\emph{i.e.}, the node having  collocated, aggregated LTE eNodeB and Wi-Fi access point functionalities) is implemented in NS-3 which introduces a traffic steering layer (\emph{i.e.}, Link Aggregation Layer) for efficient integration of LTE and Wi-Fi. Using extensive simulations, we verified the correctness of C-LWIP module in NS-3 and evaluated the aggregation benefits over standalone LTE and Wi-Fi networks with respect to varying number of users and traffic types. We found that split bearer performs equivalently to switched bearer for UDP flows and switched bearer outperforms split bearer in the case of TCP flows. Also, we have enumerated the potential challenges to be addressed for unleashing C-LWIP capabilities. Our findings also include WoD-Link Aggregation Strategy which is shown to improve system throughput by 50\% as compared to Naive-LAS in a densely populated indoor stadium environment. 

\end{abstract}
\IEEEpeerreviewmaketitle


\section{Introduction}
\label{sec:intro}
\par The penetration of multi-featured electronic gadgets such as smart phones, tablets, laptops in the market and popularity of mobile applications (native and web) developed for these devices have significantly increased the data traffic demand from mobile subscribers. According to Cisco VNI forecast smart phones generate approximately $1$ GB of data per month which is nearly $40$ times of the data generated by a feature phone~\cite{cicso_vni}. Also, mobile data traffic growth will keep increasing and reach $30.6$ Exabytes per month by $2020$ compared to $3.7$ Exabytes per month in $2015$. However, the telecommunication service providers/operators face many challenges in order to improve their cellular network capacities to match these ever increasing data demands due to low, almost flat Average Revenue Per User (ARPU) and low Return on Investment (RoI). Spectrum resource crunch and licensing requirement for operation in cellular bands further complicate the procedure to support and manage the network. 
\par Utilizing unlicensed spectrum effectively by interworking of cellular/mobile network and Wi-Fi networks is shown to be a potential candidate technology to solve the data crunch problem. Numerous interworking architectures were proposed in the literature. In~\cite{7060499}, authors presented three different architectures for realizing interworking, (1) loosely coupled, (2) tightly coupled and (3) hybrid architecture. Loosely coupled architecture of LTE and Wi-Fi is proposed for non-collocated scenario, where LTE and Wi-Fi networks are connected through P-GW. It is suggested that multipath TCP (MPTCP) can be used for realizing loosely coupled architecture, which can take intelligent decisions for traffic steering at transport layer. Tightly coupled architecture shows that LTE and Wi-Fi radios are tightly bound and there exists only one core network for both access networks. This tight interworking realizes the potential of finer control over available radio interfaces in decision making and flow routing based on the channel states. Hybrid integration suggests a tighter integration to be realized along with merits of loosely coupled architecture.

The tightly coupled architecture is chosen as a study item and standardized recently by 3GPP. The tighter integration of LTE and Wi-Fi is included as  part of Rel~13, which has the following advantages: 
\begin{itemize}
\item Wi-Fi operations are controlled directly via LTE base station (eNB) and therefore LTE core network (\emph{i.e.}, Evolved Packet Core (EPC)) need not manage Wi-Fi separately.
\item Radio level integration allows effective radio resource management across Wi-Fi and LTE links.
\item LTE acts as the licensed-anchor point for any UE, providing unified connection management with the network.
\end{itemize}
Tightly coupled architecture is observed to have a finer level of control on radio interfaces. The integration of LTE and Wi-Fi can be realized at different layers of LTE protocol stack viz., IP, PDCP, RLC and MAC layers. An architectural proposal to 3GPP for realizing tighter level of interworking at PDCP level utilizes the split bearer and switched bearer properties of dual connectivity~\cite{TS36842} to steer traffic across two radios effectively. This proposal is standardized by 3GPP as LTE Wi-Fi Aggregation (LWA)~\cite{36300}. In LWA the packets received through both interfaces are reordered at PDCP layer and delivered to higher layer in-order. The performance benefits of LWA at PDCP layer of LTE protocol stack is given in~\cite{lagrange2014very}. Another architecture proposal suggests aggregation at RLC layer~\cite{RP150180}. This supports steering of packets from LTE to Wi-Fi from the RLC buffer. RLC retransmission and reordering ensures the reliability of the flows. It is shown that aggregation at RLC layer performs better than MPTCP. The performance evaluation of RLC level interworking is given in~\cite{LWIR}.For implementing both the architectures, changes have to be made at the protocol stack of UE and eNB. This makes these architectures not suitable for existing commercially available UEs to readily use these architectures even with the availability LTE and Wi-Fi interfaces. 

3GPP has recently standardized LTE Wi-Fi interworking at IP layer with an IPSec tunnel (LWIP). The LWIP  supports the existing UEs to readily interwork without any protocol changes~\cite{LWIP_DEMO}. Also, LWIP supports switched bearer and split bearer for steering traffic across LTE and Wi-Fi links. Switched bearer switches a bearer of the UE completely from LTE interface to Wi-Fi interface, whereas split bearer splits an existing bearer across LTE and Wi-Fi interfaces. 

\begin{figure}[!htb]
\centering
\epsfig{width=4cm, figure=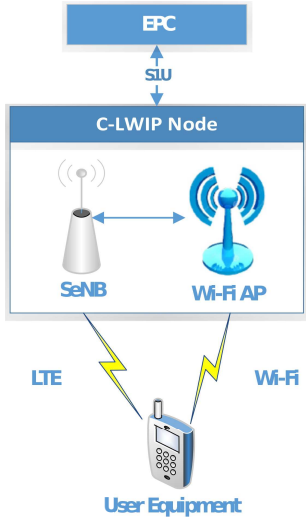}
\caption{A C-LWIP Realization.}
\vspace{-0.3cm}
\label{fig:intro}
\end{figure}
We have investigated a collocated LTE/Wi-Fi scenario (C-LWIP) in which LTE small cell eNodeB (SeNB) and Wi-Fi access point (AP) are tightly coupled at RAN level as shown in Fig.~\ref{fig:intro}.


\textbf{Contributions : }Our main contributions in this paper are,
\begin{itemize}
\item Proposed a C-LWIP architecture which mitigates interference in a dense urban deployment. This architecture ensures the interworking benefits to existing commercial UEs without any protocol changes.
\item Implemented C-LWIP module in NS-3 simulator and validated its correctness through extensive simulations.
\item Using C-LWIP module, we have evaluated the aggregation benefits for different link aggregation strategies.
\end{itemize}
\begin{figure}[!htb]
\centering
\epsfig{width=8cm, height=8cm, figure=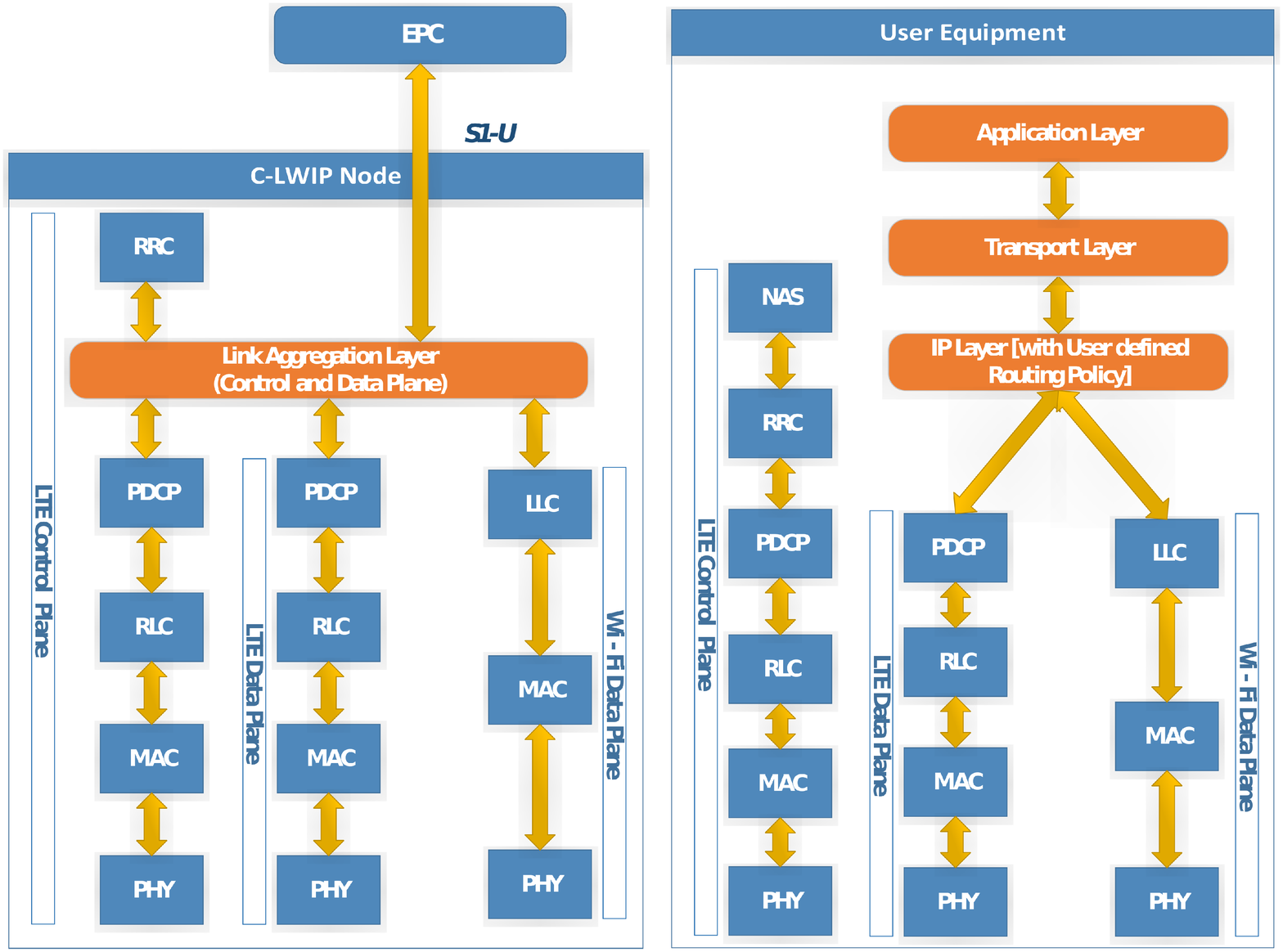}
\caption{Proposed Architecture of C-LWIP.}
\vspace{-0.6cm}
\label{fig:proposed}

\end{figure}

\section{LWIP Architectures}
\label{sec:proposedSystem}
3GPP has defined a LWIP framework by aggregation of non-collocated LTE and Wi-Fi at IP level which uses IPSec tunnel between UE and eNodeB to interwork with Wi-Fi~\cite{lwip3gpp}. We proposed C-LWIP framework using a \textit{Link Aggregation Layer} (LAL) in to Telecommunications Standards Development Society (TSDSI), India~\cite{SWIP50} (a daughter body of 3GPP). The proposed architecture makes existing UEs to readily work with C-LWIP node without any protocol modification at UE side.
\subsection{LWIP Architectures Overview}
\subsubsection{\textbf{Non-collocated LWIP Architecture}} 
The architecture is shown in~\cite{git}. The eNodeB utilizes the available Wi-Fi radio resources for the UEs in RRC\_CONNECTED state. A secure IP tunnel is established between LWIP node and UE to ensure security for communication over Wi-Fi interface. This IPsec tunnel has increased the overhead of communication in a collocated scenario compared to our proposed architecture. In a non-collocated scenario, IPsec tunnel holds true. The decision to steer traffic in LTE or Wi-Fi link is communicated to UE via a specific \textit{steering command} which in turn is notified to the higher layers on the protocol stack. The higher layers can also take decision for traffic steering to Wi-Fi interface.
\begin{figure}[htb]
\centering
\epsfig{width=5cm,figure=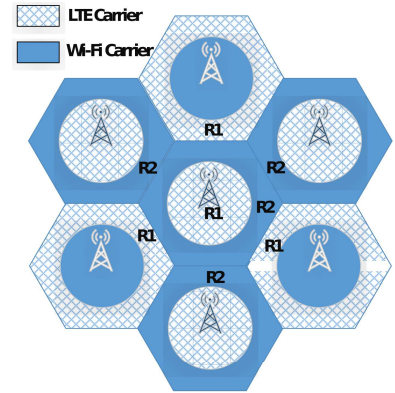}
\caption{Frequency Reuse in a dense C-LWIP deployment}
\label{fig:rem}
\end{figure}
\subsubsection{\textbf{Collocated LWIP Architecture}} 
In this proposed architecture, C-LWIP is realized by introducing a \textit{Link Aggregation Layer} (LAL) in the protocol stack of C-LWIP node as shown in Fig.~\ref{fig:proposed}. This makes even the existing UEs to readily benefit from C-LWIP without any modifications. LAL does not add any new headers to the IP data packets received from EPC via the S1-U interface. Packets going through LTE and Wi-Fi interfaces follow regular packet forwarding procedures at their protocol stacks and get delivered directly to IP layer. LAL can collect various network parameters and actively participate in intelligent decision making for steering IP traffic across LTE and Wi-Fi interfaces in the downlink. The security over Wi-Fi interface is provided by Wi-Fi key, which is obtained from LTE RRC and communicated to UE through RRC signaling. RRC generates the Wi-Fi Layer 2 security key from eNB Key $K_{eNB}$. In our proposed C-LWIP architecture, the traffic split can be realized both at packet and flow level. In flow level split, an entire flow is sent on either LTE or Wi-Fi link and hence in-sequence delivery issue does not arise. In packet level split, individual packets of a flow can be sent across different interfaces, hence, achieving in-sequence delivery in such a split is a challenging problem.
\subsubsection{\textbf{Advantages of Collocated over Non-collocated LWIP Architecture}} 
The tighter level of aggregation between LTE eNodeB and Wi-Fi AP in collocated fashion in C-LWIP node has several advantages with respect to end user throughput in contrast to the non-collocated architecture, both at the physical layer and network layer.
\textbf{\\At Physical Layer:}
The major advantage of collocated architecture as compared to non-collocated architectures is the flexibility in adapting fractional frequency reuse (FFR) scheme for mitigating inter-cell interference. Given a spatial distribution of UEs in the coverage, C-LWIP node may employ FFR where LTE eNodeB of C-LWIP serves users in the inner region and Wi-Fi of C-LWIP node serves the interference-prone LTE cell-edge users. In case of dense urban scenarios, C-LWIP nodes act as an important contributor for mitigating the interference among neighbor C-LWIP nodes by assigning non-overlapping LTE and Wi-Fi bands appropriately as shown in Figure~\ref{fig:rem}. Multiple C-LWIP nodes are considered whose coverage regions are spatially marked distinctly as regions $R1$ and $R2$. If the users exist in $R1$, they will be served by LTE interface of C-LWIP. Similarly, if a user resides in region $R2$, then it could potentially use Wi-Fi AP to serve the users to mitigate the inter-cell interference. This is possible due to unified control plane signaling between LTE eNodeB and Wi-Fi AP in C-LWIP node.

In the case of non-collocated architecture, the interference mitigation  could not be achieved effectively when LTE and Wi-Fi radios are placed far apart. This is because controlling transmit power of LTE or Wi-Fi link may lead some user to go out of coverage region. This prevents us from employing FFR scheme to non-collocated LWIP. Hence, only data plane offloading of LTE traffic to Wi-Fi can be supported.

\textbf{At Network Layer:} 
IPSec tunnel introduced in non-collocated deployment involves encryption of packets at IP layer (to send through untrusted WLAN terminal) followed by link level encryption of WLAN (optionally) which can be removed in a collocated scenario. Our proposed architecture reduces the overhead of double encryption (\emph{i.e.}, at IP and Layer 2 of WLAN) by using Wi-Fi key per client derived from existing eNB key $K_{eNB}$. Also, every packet sent through IPSec tunnel is added with tunnel endpoint header, which adds to inefficient use of resources over the wireless channel. Collocated architecture does not require any additional headers. Non-collocated architecture facilitates to readily work with existing Wi-Fi AP, but the decision for traffic offloading is simplified at a coarse level of granularity \emph{e.g.}, observed throughput and delay over an interface can be the determining factor for taking the offloading decision. But collocated architecture supports decision making for offloading at a very fine granularity of information \emph{i.e.}, channel load, received SNR of Wi-Fi and channel characteristics such as loss and fading. This helps collocated architecture to perform better than non-collocated architecture.

\subsection{Link Aggregation Strategies}
\label{subsec:comparison}

In this subsection, we present two link aggregation strategies (LASs) for C-LWIP. 
\begin{enumerate}
\item \textit{\textbf{Naive LAS or N-LAS}}: In this approach, LTE and Wi-Fi links are simultaneously used for sending uplink and downlink IP data traffic. In general, nearly half of the traffic is sent through LTE link and remaining half passes via Wi-Fi link irrespective of their channel conditions. It has two variants depending on whether the split is performed at packet level or flow level.
\begin{itemize}                                                                                                                                                                                                                                                                                                \item  Packet Split N-LAS : Split equally within a single IP flow.
\item  Flow Split N-LAS : Split equally among multiple IP flows but a flow is routed via one of the links.
\end{itemize}
\item \textit{\textbf{Wi-Fi only on Downlink LAS or WoD-LAS}}: In this approach, Wi-Fi is used for transmitting downlink traffic while LTE is used for transmitting both uplink and downlink traffic as shown in~\cite{git}. As the number of users increases in the network, due to CSMA/CA, contention in Wi-Fi network also increases which brings down the throughput of Wi-Fi network. WoD-LAS was proposed in~\cite{7060499} and presented as a tightly coupled interworking architecture. WoD-LAS lowers the possibility of contentions in Wi-Fi link as it involves only downlink transmissions.
\end{enumerate}

\section{C-LWIP Module in NS-3}
\label{sec:lwans3}
We developed a C-LWIP module in NS-3~\cite{ns3} to evaluate the performance of different link aggregation strategies. An essential component of this design is to realize C-LWIP node, which is achieved by binding the LTE and Wi-Fi radio interfaces together at IP level. This binding is implemented by a class known as \textit{LinkAggregationLayer}. This class is responsible for provisioning various dynamic link level schedulers and steering algorithms. A high-level class diagram of C-LWIP design is given in~\cite{git}. 

When a packet is received from LTE core network (via S1-U) at C-LWIP node, LTE specific packet headers (GTP headers) are removed and resulting packet is routed to appropriate radio interface (\emph{i.e.}, LTE or Wi-Fi \textit{netdevices}) dictated by the LAS. A map for MAC address of the UE to IP address of the UE is created. When a new packet with destined IP address arrives to be sent via Wi-Fi, it is placed into LLC of Wi-Fi with the help of destination MAC address obtained from the map. An entry is made in this mapping table when a UE associates with the C-LWIP node. To send a packet through LTE interface, the packet is forwarded to LTE socket at C-LWIP node. 
To send a packet through Wi-Fi interface, UE's Wi-Fi MAC address is retrieved by using a mapping table at \textit{LinkAggregationLayer} in the C-LWIP node.

A network address translation mechanism is devised at UE side in order to route traffic via unified connection management. This is driven by the fact that, LTE works as the anchor point for Wi-Fi node and no route exists between Wi-Fi and public Internet other than through LTE EPC. At UE side, packets generated by an application are routed to any of the available interfaces as dictated by the link aggregation strategies. In this work, we have used LASs at the UE side as well which can be implemented by operator defined policies. Provisions are made to implement uplink steering algorithms across radio interfaces. For flow level traffic steering, a five-tuple structure is designed to create mapping for radio interface which is necessary for pushing traffic as per decisions are taken by link aggregation strategy. 

\section{Experimental Setup}
\label{sec:simResults}
The experimental platform is based on the C-LWIP module developed by extending NS-3 simulator. The simulation parameters are given in Table \ref{tab:config}. In order to simulate the scenarios realistically, we have considered the backhaul delay as 40~ms. Our simulation test bench evaluates various schemes which are described as follows.
\begin{itemize}
\item \textbf{LTE NoLAS :} All traffic between UE(s) and C-LWIP nodes is sent through LTE links.
\item \textbf{Wi-Fi NoLAS :} All traffic between UE(s) and C-LWIP nodes is sent through Wi-Fi links.
\item \textbf{Packet Split N-LAS (PS-N-LAS)}
\item \textbf{Flow Split N-LAS (FS-N-LAS)} 
\item \textbf{WoD-LAS :} Unlike FS-N-LAS, in this strategy, Wi-Fi is used only in downlink for carrying flows whereas LTE is used for both uplink and downlink. \modify{All uplink flows of UE through LTE interface is achieved by inserting appropriate forwarding rules in UE's \textit{iptable}} without any protocol stack modification.
\end{itemize}

\begin{table}[b]
\centering
\vspace{-0.4cm}
\caption{NS-3 Simulation Parameters}
\renewcommand{\arraystretch}{1.2}
\scriptsize
\begin{tabular}{|l|l|} \hline 
\label{tab:config}
\textbf{\hspace{1cm}Parameter} & \textbf{\hspace{1cm}Value}\\ [0.7ex]\hline \hline
Number of C-LWIP Nodes & 1 and 10 \\ \hline
LTE Configuration  & 10MHz, 50 RBs, FDD \\ \hline
Wi-Fi Configuration & IEEE 802.11a, 20 MHz \\ \hline
Traffic Type & Mixed (voice, video, web, FTP) \\ \hline
Distance b/w UE \& C-LWIP node & 25 Meters\\ \hline
Simulation Time & 100 Seconds\\ \hline
Error Rate Model & NIST Error Rate Model\\ \hline
Mobility Model & Static \\ \hline
Wi-Fi Rate Control Algorithm & Adaptive Auto Rate Fallback\\ \hline
LTE MAC Scheduler & Proportional Fair Scheduler\\ \hline
Number of seeds & 5\\ \hline
Wi-Fi Queue size & 400 packets\\ \hline
backhaul Delay & 40 ms\\ \hline
\end{tabular}
\end{table}

\begin{table}[h]
\centering
\vspace{1mm}
\caption{Percentage Distribution of User Traffic }
\renewcommand{\arraystretch}{1.2}
\scriptsize
\begin{tabular}{|c|c|c|c|c|} \hline 
\label{tab:mixedtraffic}
\textbf{Traffic Class} & \textbf{Nature} & \textbf{Expt \#3} & \textbf{Expt \#4} & \textbf{Expt \#5}\\ [1ex]\hline \hline
Voice & UDP & 20\% & 20\% & 40\% \\ \hline
FTP & TCP & 20\% & 60\% & 50\% \\ \hline
Video & UDP & 60\% & 20\% & 30\% \\ \hline
Web & TCP & 20\% & 40\% & 60\% \\ \hline
\end{tabular}
\end{table}
Depending on the number of C-LWIP nodes, number of UEs and nature of traffic, we have conducted five sets of experiments with different link aggregation strategies. First, two experiments (\#1 and \#2) are performed to benchmark C-LWIP benefits for an ideal case of one and four users with UDP traffic, respectively. Next two experiments (\#3 and \#4) are conducted to see the performance of C-LWIP in a typical home scenario with mixed traffic (\emph{i.e.}, voice, video, web, FTP). The last experiment (\#5) imitates a real-world indoor stadium scenario having multiple C-LWIP nodes with mixed traffic. The exact percentage of users in each of the traffic types in a mixed traffic scenario are shown in Table \ref{tab:mixedtraffic}. The details of each experiment conducted are briefly summarized as follows. 
\begin{itemize}
\item \textbf{Expt \#1 :} This experiment involves one C-LWIP node with only one user to study the ideal behaviour of the system. We considered default bearer with four UDP data flows (two in uplink and two in downlink) and observed network throughput w.r.t. UDP Application Data Rate (ADR) by varying the data rate as 1, 6, 12, 24~Mbps per flow.
\item \textbf{Expt \#2 :} It involves one C-LWIP node with four users. We considered default bearer with four UDP data flows per user (two in uplink and two in downlink), thus, with 16 flows in total for study. The network throughput is observed w.r.t. ADR by varying the data rate as 1, 2, 4, 8~Mbps per flow.
\item \textbf{Expt \#3 :} To demonstrate the interworking benefits in a typical home scenario, this experiment involves one C-LWIP node with varying number of users from five users to 30 users. It considers mixed traffic scenario having the majority of UDP flows (UDP-Heavy). 
\item \textbf{Expt \#4 :} This experiment involves one C-LWIP node with varying number of users from five users to 30 users. Unlike previous experiment, it considers mixed traffic scenario having majority of TCP flows (TCP-Heavy).
\item \textbf{Expt \#5 :} To observe the performance of C-LWIP in a real-world indoor stadium, this experiment involves 10 C-LWIP nodes with varying number of users from 50 to 400. LTE of C-LWIP node is operating with reuse factor one, and every Wi-Fi AP of C-LWIP node operates in the same channel. Realization of indoor stadium includes multiple C-LWIP nodes with diverse data traffic requirements.
\end{itemize}

\begin{figure*}[htb!]
\minipage{0.32\textwidth}
\epsfig{width=5.7cm,height=4cm,figure=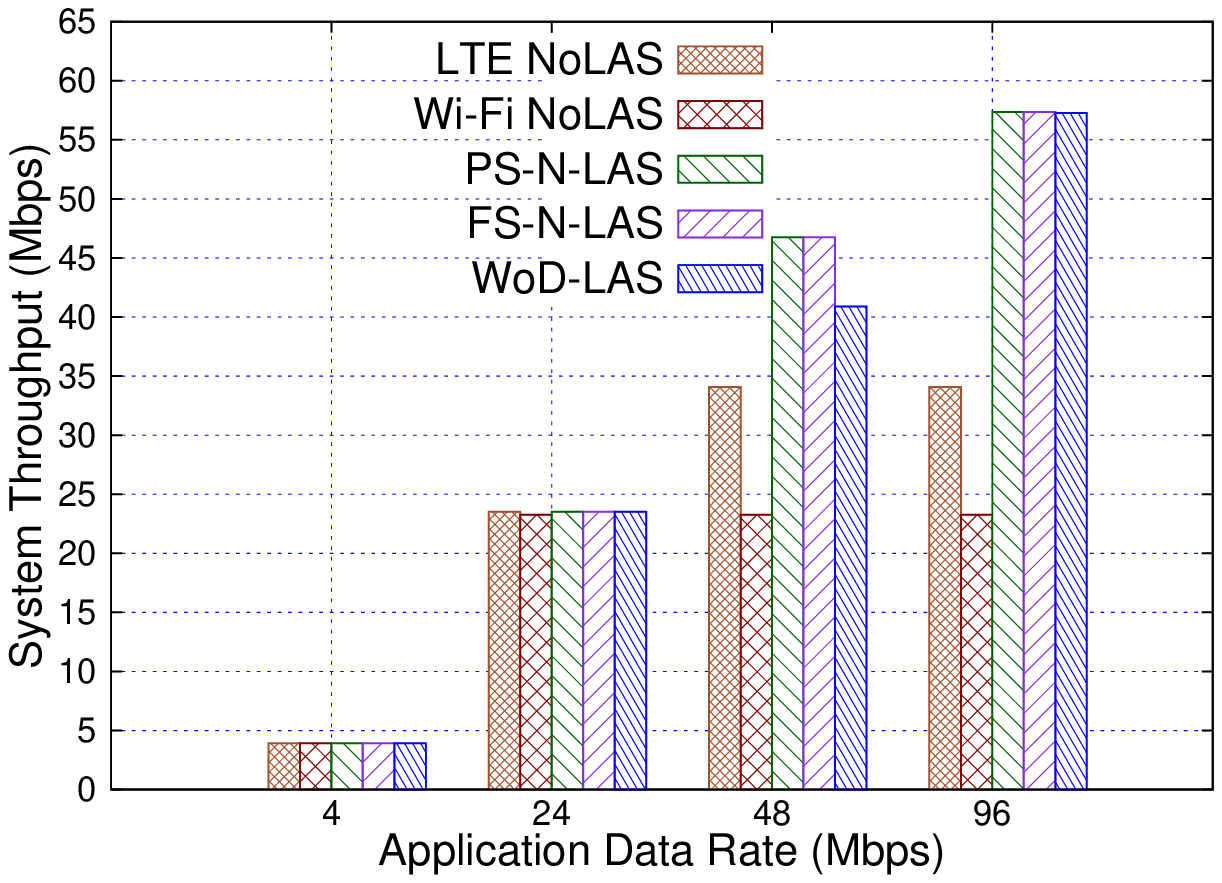}
 \vspace{-0.17cm}
\caption{UDP Network Throughput : One UE}
\label{fig:udp1uethroughput}
    \vspace{-0.3cm}
\endminipage\hfill
~
\minipage{0.32\textwidth}
\epsfig{width=5.7cm,height=4cm,figure=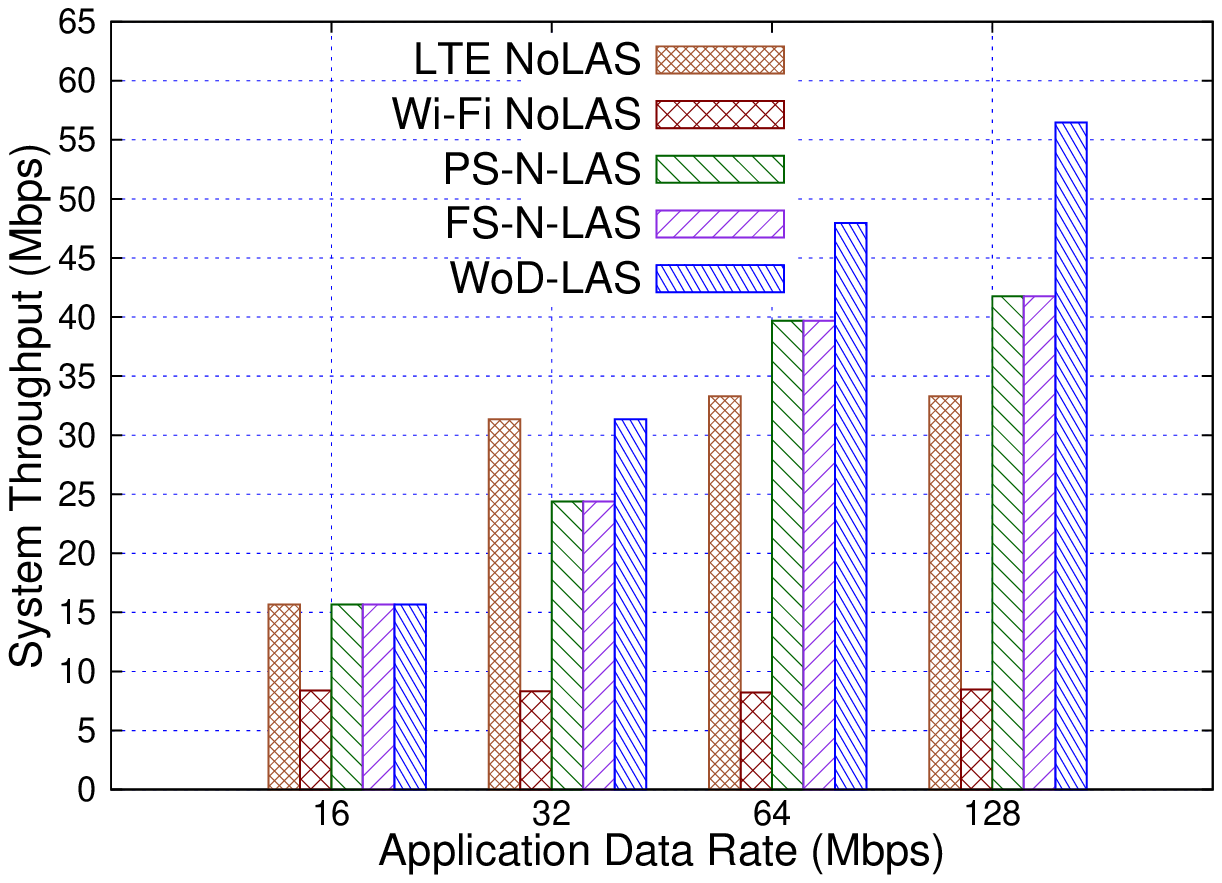}
 \vspace{-0.17cm}
\caption{UDP Network Throughput : Four UEs}
\label{fig:udp4uethroughput}
    \vspace{-0.3cm}
\endminipage\hfill
~
\minipage{0.32\textwidth}
\epsfig{width=5.7cm,height=4cm,figure=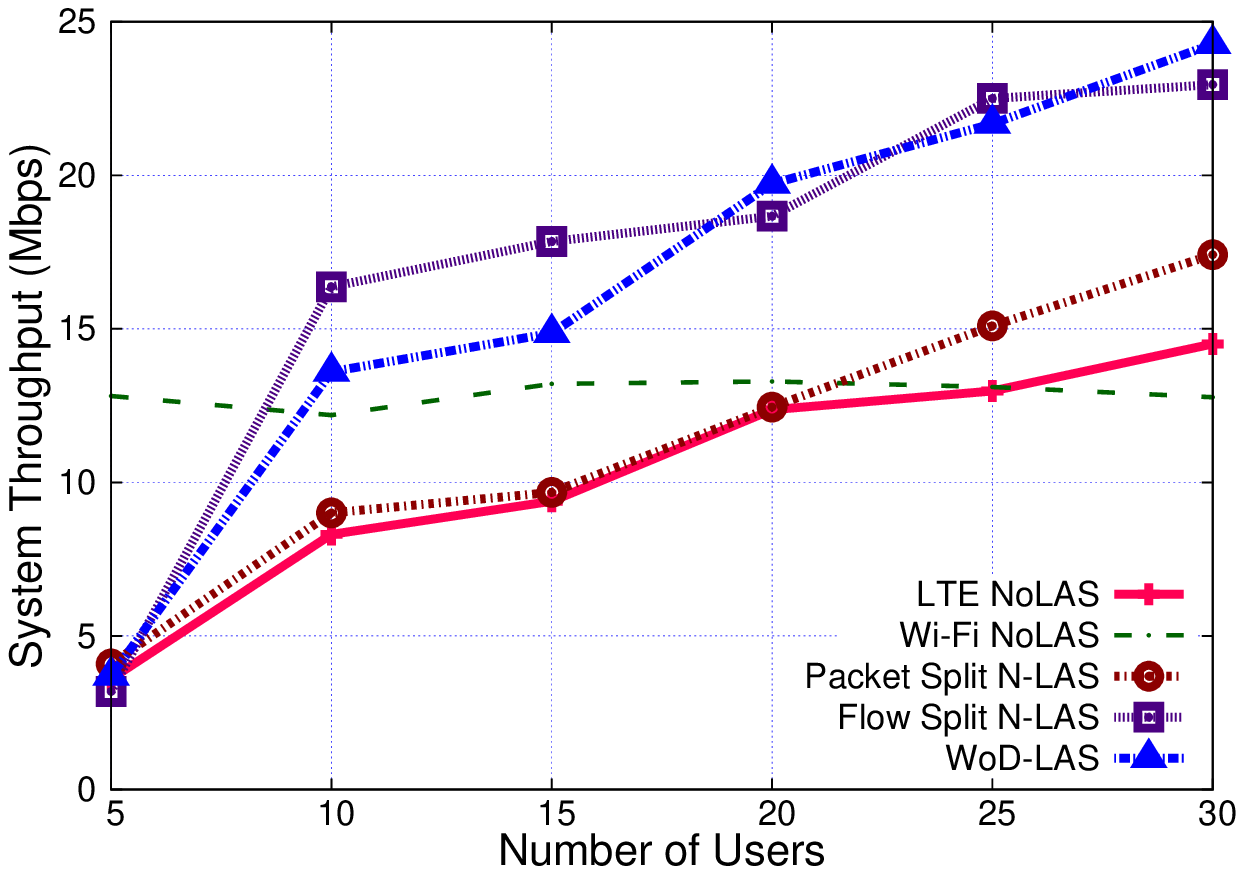}
 \vspace{-0.17cm}
\caption{Home Scenario with one C-LWIP Node, Mixed Traffic, UDP Heavy}
\label{fig:home_udp}
    \vspace{-0.3cm}
\endminipage\hfill
\end{figure*}
\begin{figure*}[htb!]
\minipage{0.32\textwidth}
\epsfig{width=5.7cm,height=4cm,figure=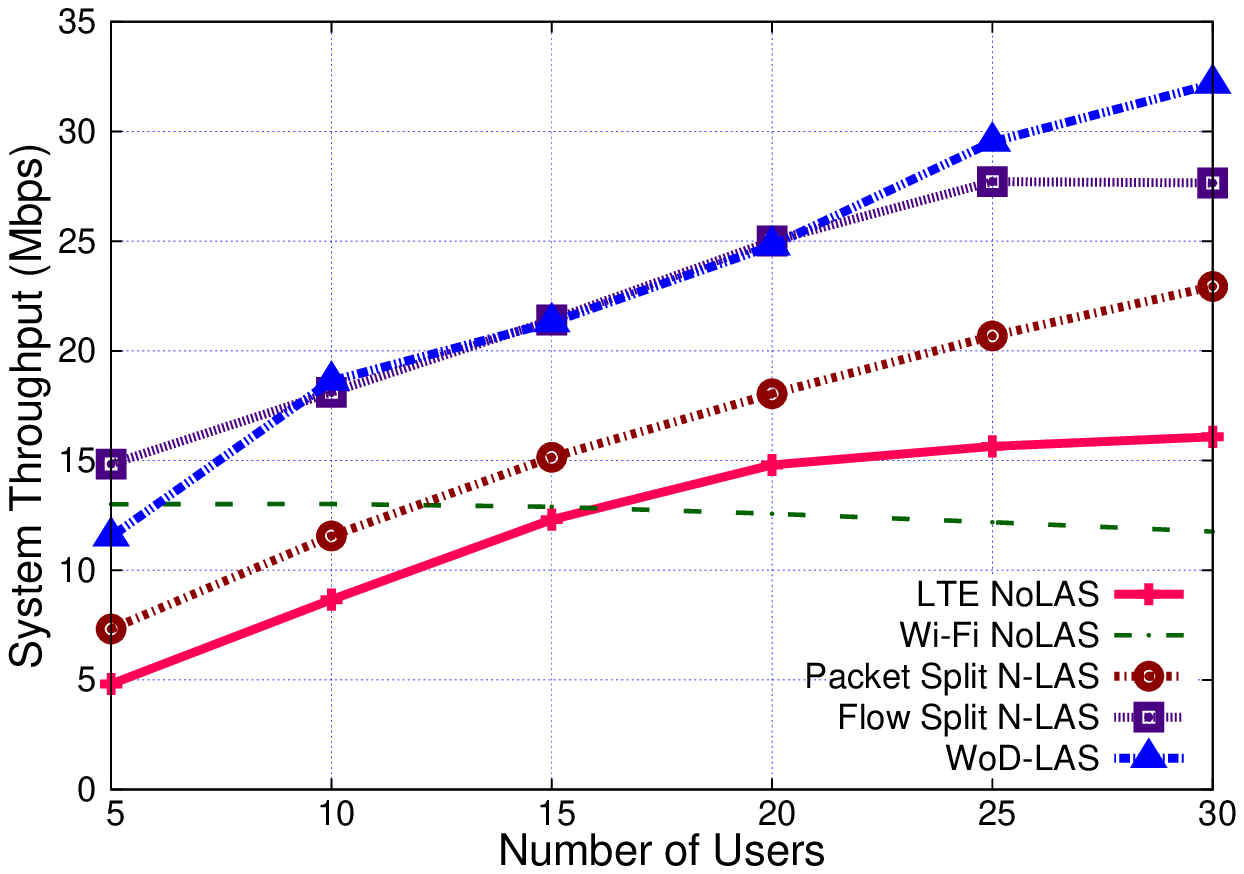}
 \vspace{-0.17cm}
\caption{Home Scenario with one C-LWIP Node, Mixed Traffic, TCP Heavy}
\label{fig:home_tcp}
    \vspace{-0.3cm}
\endminipage\hfill
~
\minipage{0.32\textwidth}
\epsfig{width=5.7cm,height=4cm,figure=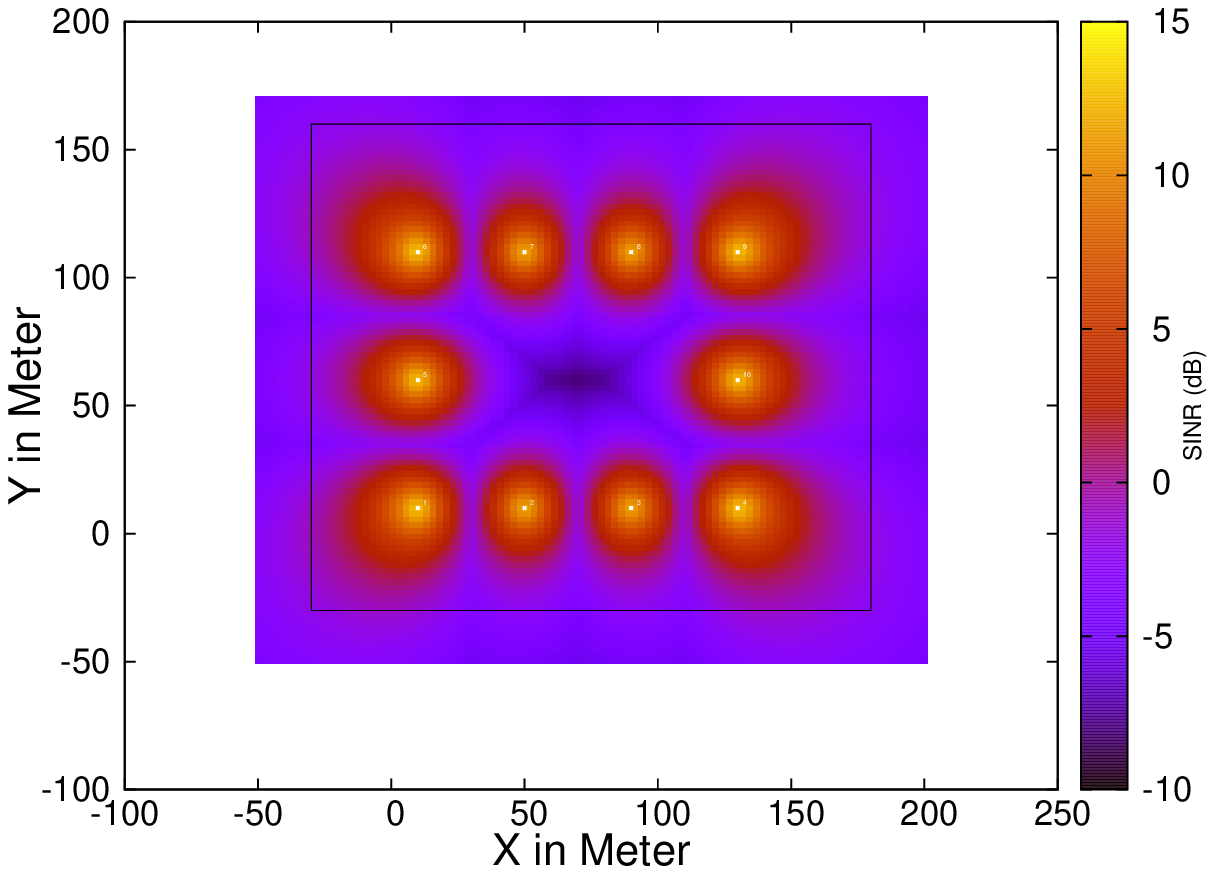}
 \vspace{-0.17cm}
\caption{REM Plot for Indoor Stadium layout with 10 C-LWIP Nodes}
\label{fig:remplot}
    \vspace{-0.3cm}
\endminipage\hfill
~
\minipage{0.32\textwidth}
\epsfig{width=5.7cm,height=4cm,figure=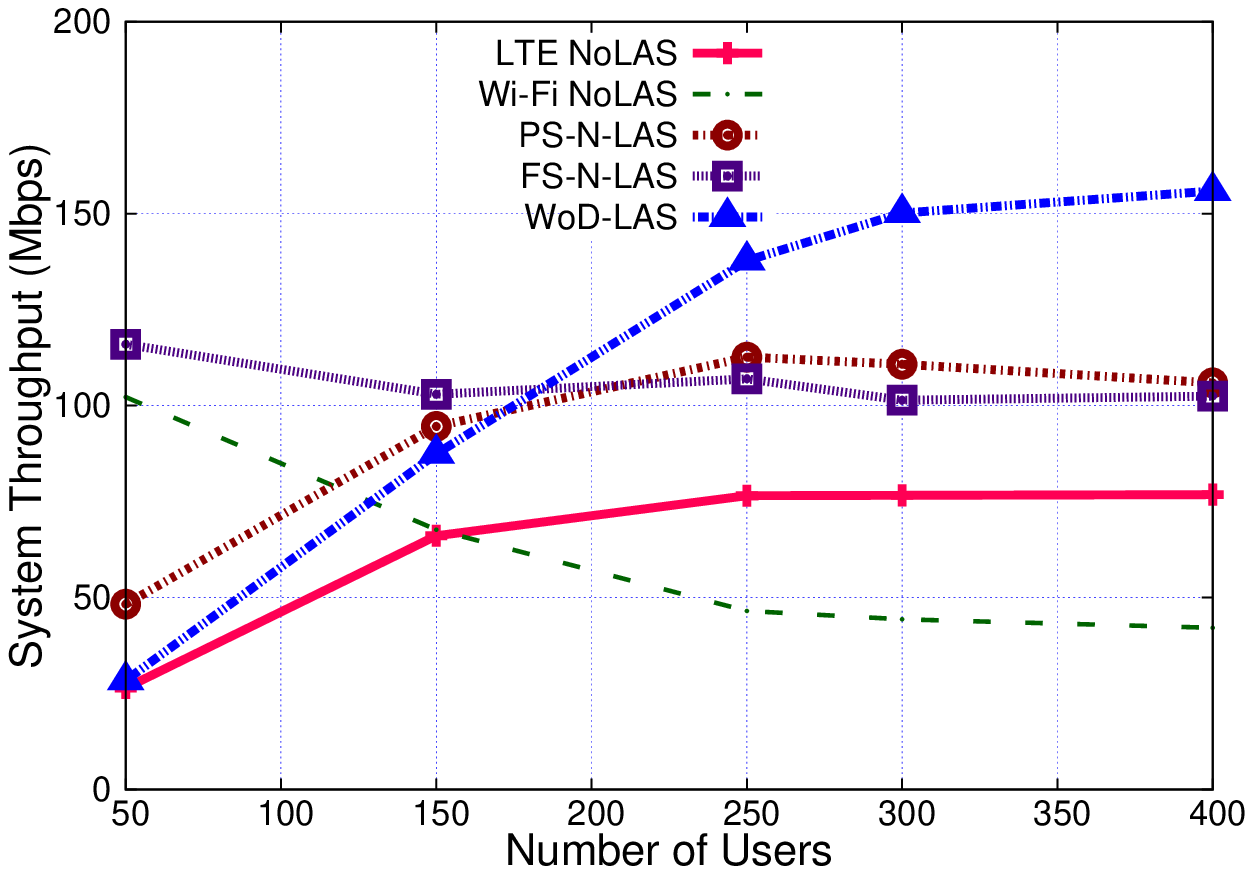}
 \vspace{-0.17cm}
\caption{Indoor Stadium with 10 C-LWIP Nodes, Mixed Traffic}
\label{fig:indoorstadium}
    \vspace{-0.3cm}
\endminipage\hfill
\end{figure*}

\section{Performance Results}
\label{sec:perfAnalysis}
The variation of UDP throughput w.r.t UDP ADR of uplink and downlink flows for one UE and four UEs are shown in Figs.~\ref{fig:udp1uethroughput} and~\ref{fig:udp4uethroughput}, respectively. UDP traffic types tend to harvest maximum capacity of the links, hence this experimental result sets a classical benchmark for aggregation advantages over individual LTE and Wi-Fi radio links. 
\subsection{Analysis of Expt \#1}In one UE case with 4~Mbps and 24~Mbps ADR, the network is able to deliver the data in all the LASs (as shown in Fig.~\ref{fig:udp1uethroughput}). The throughput variation in Wi-Fi NoLAS does not vary much after 48~Mbps ADR and thereafter saturates, because, it reaches its maximum achievable rate of 24~Mbps for 802.11a with maximum PHY rate of 54~Mbps. Similarly, LTE NoLAS attains saturation after 48~Mbps. However, leveraging the interworking benefits of C-LWIP node, PS-N-LAS and FS-N-LAS are able to deliver higher network throughputs than that of individual LTE and Wi-Fi only networks. The two variants of N-LAS schemes are indistinguishable in performance due to its naive approach of equally dividing flows and type of user traffic. WoD-LAS is no better than both FS-N-LAS and PS-N-LAS due to presence of only one user and no contention in Wi-Fi link. The next experiment encompasses a contention based scenario. 
\subsection{Analysis of Expt \#2}The inclusion of four users in the network leads to contentions and therefore, Wi-Fi only performance is observed to be poor as compared to other LASs. As LTE operates on the principle of scheduler based MAC, its throughput continues to rise with an increase in ADR but attains saturation after 34~Mbps (as shown in Fig.~\ref{fig:udp4uethroughput}). Like the previous experiment, this experiment shows almost equal throughputs due to naive approach of equally dividing flows and type of user traffic. 
An important takeaway by comparing the results between N-LAS and WoD-LAS is that contentions of Wi-Fi degrades the N-LAS performance resulting lower peak value than WoD-LAS throughput. However, WoD-LAS does not suffer from this drawback by preventing contentions in Wi-Fi, as Wi-Fi link is used only in downlink.
\subsection{Analysis of Expts \#3 and \#4}In order to understand the behavior of C-LWIP node for a typical home deployment scenario, the next two experiments demonstrate performance benefits of C-LWIP considering UDP heavy and TCP heavy mixed traffic which are shown in Figs.~\ref{fig:home_udp} and~\ref{fig:home_tcp}, respectively. In both plots, with increase in number of users, the aggregation of LTE and Wi-Fi has resulted enhanced throughput than LTE NoLAS and Wi-Fi NoLAS. Wi-Fi performance is degraded due to the occurrence of contentions and does not improve. Packet split mechanism could not improve proportionally due to inherent issue of out-of-order deliveries and Dupack transmissions for TCP flows. These problems are avoided in flow split N-LAS, because, a  flow is pushed as a unit to destined radio interface. Comparison of WoD-LAS and FS-N-LAS shows that WoD-LAS suppresses the demerits of FS-N-LAS by restricting the downlink user flows to Wi-Fi and LTE, and uplink flows only to LTE. In WoD-LAS, Wi-Fi utilizes its spectrum resources to carry user data and provides the best effort services by smartly utilizing the flow constraints in one direction. This facilitates a significant reduction in number of collisions, thereby improving the system throughput over N-LAS schemes.
\subsection{Analysis of Expt \#5} The variation of system throughput with large number of UEs in an indoor stadium scenario having 10 C-LWIP nodes is shown in Fig.~\ref{fig:indoorstadium}. Fig.~\ref{fig:remplot} shows the LTE Radio Environment Map (REM) of the indoor stadium layout. Clearly, like the previous experiment, Wi-Fi performance degradation is largely contributed by collisions. On the other hand, LTE throughput tends to produce less and nearly flat variation, because available radio resources are shared among all the active users. PS-N-LAS and FS-N-LAS do not show a notable difference as both the schemes are largely affected by reduced throughput on Wi-Fi link. WoD-LAS results in higher system throughput over all other LAS under study. WoD-LAS achieves a system throughput of 155~Mbps for 400 users in Fig.~\ref{fig:indoorstadium} and shows nearly 50\% more throughput than that of two variants of N-LAS.  
\begin{figure*}[t]
\minipage{0.32\textwidth}
\epsfig{width=5.7cm,height=4cm,figure=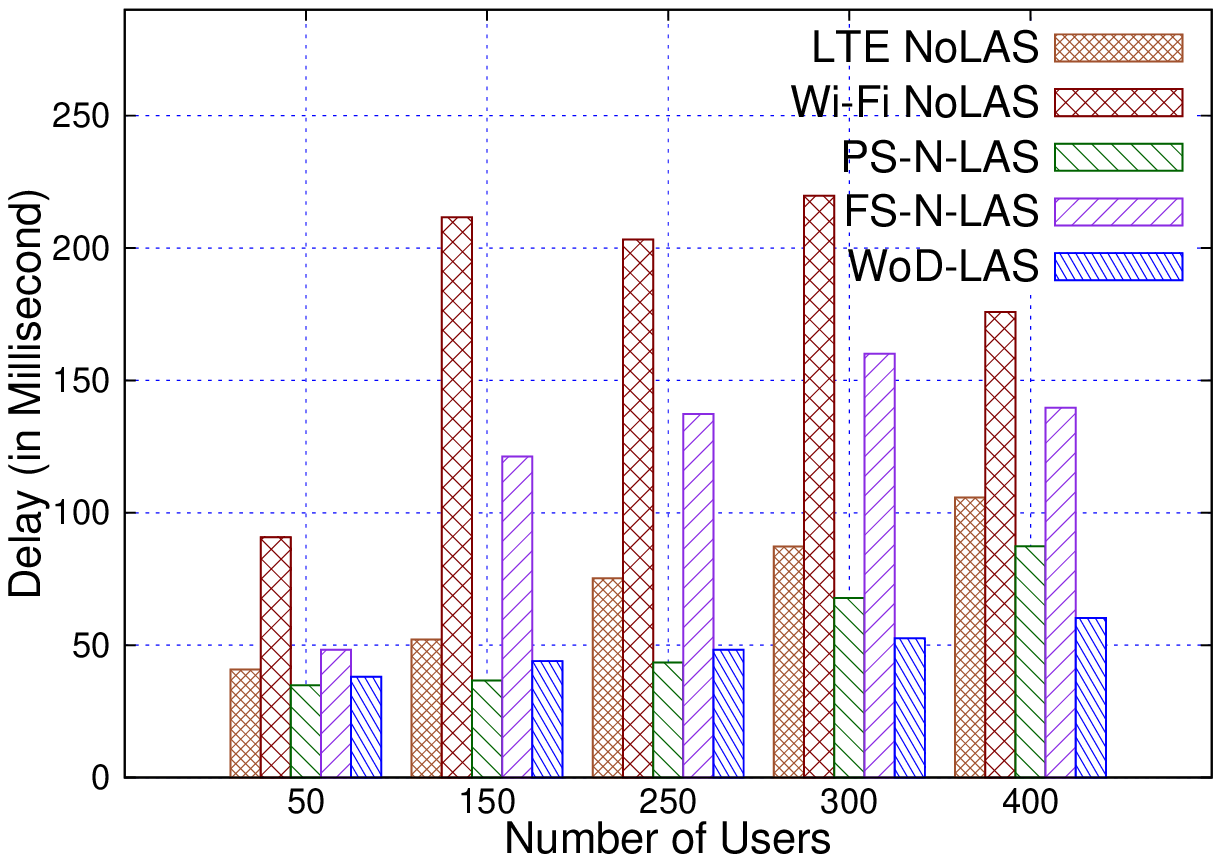}
 \vspace{-0.17cm}
\caption{Delay of Voice Traffic in Indoor Stadium with 10 C-LWIP Nodes}
\label{fig:stadlwadelay}
    \vspace{-0.3cm}
\endminipage\hfill
~
\minipage{0.32\textwidth}
\epsfig{width=5.7cm,height=4cm,figure=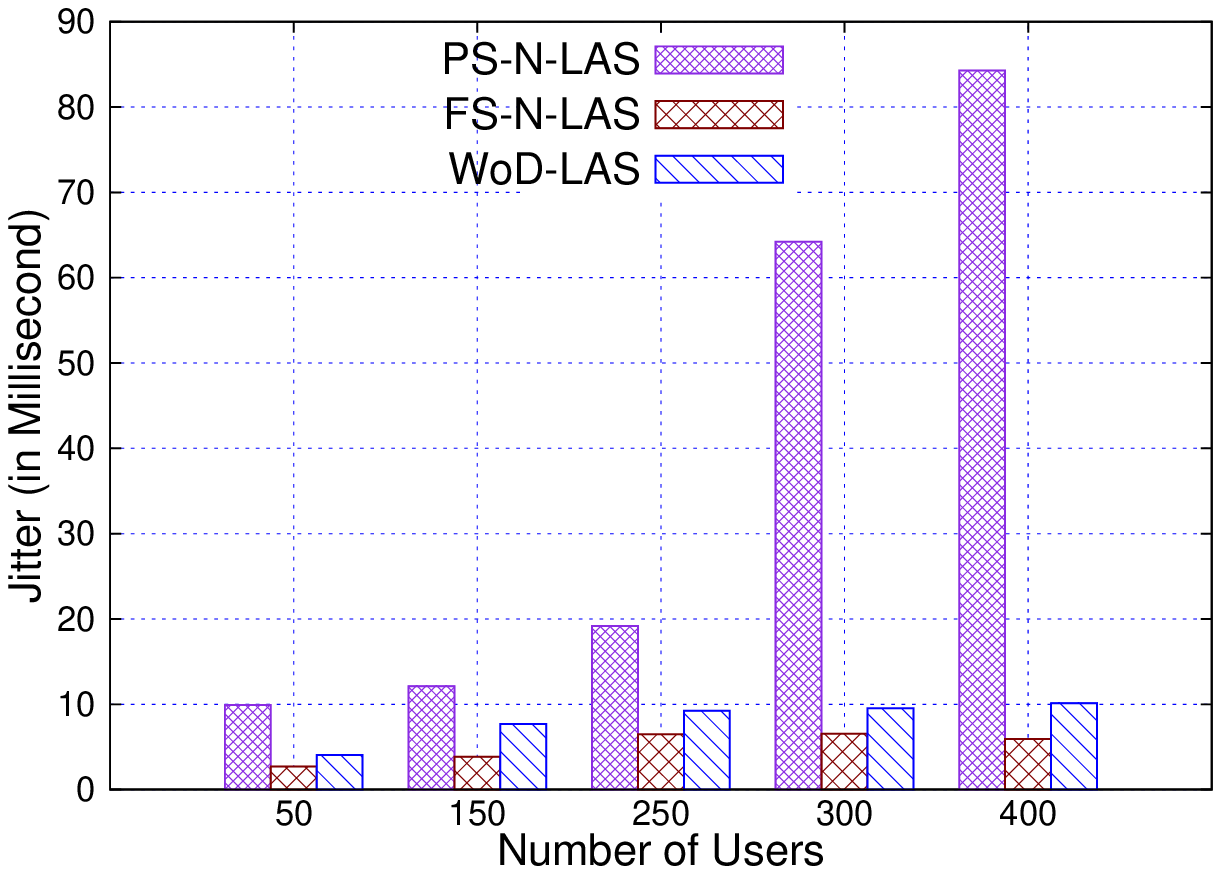}
 \vspace{-0.17cm}
\caption{Jitter of Voice Traffic in Indoor Stadium with 10 C-LWIP Nodes}
\label{fig:stadlwajitter}
    \vspace{-0.3cm}
\endminipage\hfill
~
\minipage{0.32\textwidth}
\epsfig{width=5.7cm,height=4cm,figure=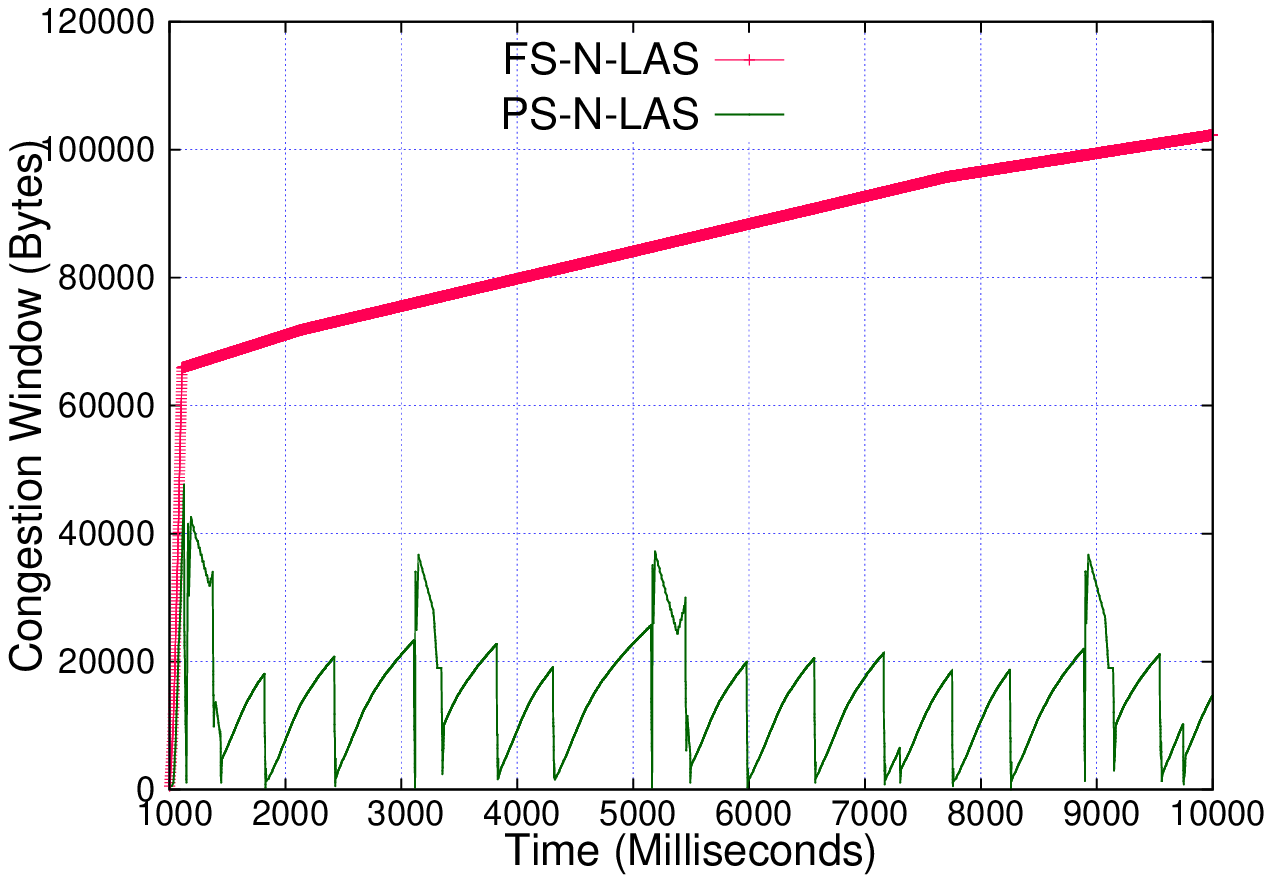}
 \vspace{-0.17cm}
\caption{Variation of TCP Congestion Window for FTP traffic flows}
\label{fig:lwacw}
    \vspace{-0.3cm}
\endminipage\hfill
\end{figure*}
In Fig.~\ref{fig:stadlwadelay}, PS-N-LAS experiences less end-to-end delay as compared to other LASs, because of two radio interfaces with different packet service rates. FS-N-LAS incurs higher delay than PS-N-LAS as all packets within the flow are routed to one interface. With less number of users, WoD-LAS delay is higher than PS-N-LAS, but for large number of users, Wi-Fi contention plays a role, thus increasing delay of PS-N-LAS in case of 300 and 400 users as compared to WoD-LAS. Fig.~\ref{fig:stadlwajitter} shows variation in jitter trend for three link aggregation strategies, where PS-N-LAS has maximum jitter because Wi-Fi and LTE support different PHY data rates for their packet transmission. The jitter for FS-N-LAS and WoD-LAS are much less than that of PS-N-LAS and does not significantly impact voice traffic. Depending on network requirements, operators could dynamically switch among available LAS for enhancing user experience and responsiveness of the system. Fig.~\ref{fig:lwacw} reveals that congestion window growth of FS-N-LAS is better as compared to PS-N-LAS. This is due to out-of-order delivery of TCP packets at receiver side in LWIP architecture. A best offloading algorithm ensuring minimal out-of-order delivery will be studied as part of future work.
\section{Challenges and Solutions for C-LWIP}
\subsection{Packet split (split bearer) is advantageous:} Packet split has a finer offloading granularity and finds its goodness in the case of UDP transmissions. It is an ideal offloading solution, where offloading decisions are instantaneous based on interface availability and loss at that particular interface. A finer level of offloading is also beneficial only if the interface information is available. Current 3GPP specification on C-LWIP do not yield a better packet level steering solution, because of limited knowledge on Wi-Fi for decision making. Complete knowledge of an interface is required for an efficient decision. This makes C-LWIP as a natural choice over non-collocated LWIP.
\subsection{Packet split is also a killer strategy for TCP growth:} Even supporting dynamic offloading mechanism and finest offloading granularity, packet split is not able to offer better throughput because of difference in time of delivery of the packets to the destination. TCP is a highly reliable protocol on observing a missing packet (which is due to delay in another interface - but not lost) starts retransmission procedure by sending duplicate acknowledgements (DUPACK). TCP sender understands these DUPACKs as actual packet loss due to congestion in the network and reduces the congestion window on receiving three consecutive DUPACKs, which is the most undesirable reaction. This problem arises because IP layer fails to reorder the packets which are received out-of-order. A reordering mechanism to ensure in-order deliver of packets in case of split bearer mechanism is needed for reaping in full benefit of packet split in C-LWIP.
%
%

\section{Conclusions and Future Work}
\label{sec:conclusion}

In this paper, we have proposed a C-LWIP architecture and enumerated its benefits over 3GPP LWIP architecture. The proposed C-LWIP architecture is carefully developed such that it does not impose any protocol level modification at UE side and makes the existing commercial UE to readily work with C-LWIP. We developed a C-LWIP module by extending NS-3 simulator which serves as an experimental platform to  evaluate the performance of C-LWIP architecture. The simulation workbench supports various existing traffic steering schemes and capable of handling the design of intelligent traffic steering algorithms. It is shown that 50\% improvement in system throughput is observed for WoD-LAS, as compared to N-LAS in an indoor stadium environment. 

\section*{ACKNOWLEDGEMENT}\label{ack}
This work was supported by the project "Converged Cloud Communication Technologies", Meity, Govt. of India.

\bibliographystyle{IEEEtran}
\bibliography{references}

\begin{thebibliography}{10}
\providecommand{\url}[1]{#1}
\csname url@samestyle\endcsname
\providecommand{\newblock}{\relax}
\providecommand{\bibinfo}[2]{#2}
\providecommand{\BIBentrySTDinterwordspacing}{\spaceskip=0pt\relax}
\providecommand{\BIBentryALTinterwordstretchfactor}{4}
\providecommand{\BIBentryALTinterwordspacing}{\spaceskip=\fontdimen2\font plus
\BIBentryALTinterwordstretchfactor\fontdimen3\font minus
  \fontdimen4\font\relax}
\providecommand{\BIBforeignlanguage}[2]{{%
\expandafter\ifx\csname l@#1\endcsname\relax
\typeout{** WARNING: IEEEtran.bst: No hyphenation pattern has been}%
\typeout{** loaded for the language `#1'. Using the pattern for}%
\typeout{** the default language instead.}%
\else
\language=\csname l@#1\endcsname
\fi
#2}}
\providecommand{\BIBdecl}{\relax}
\BIBdecl

\bibitem{cicso_vni}
\BIBentryALTinterwordspacing
Cisco. Global mobile data traffic forecast update, 2015 to 2020 white paper.
  [Online]. Available: \url{http://www.cisco.com}
\BIBentrySTDinterwordspacing

\bibitem{7060499}
J.~Ling, S.~Kanugovi, S.~Vasudevan, and A.~Pramod, ``Enhanced capacity and
  coverage by wi-fi lte integration,'' \emph{IEEE Communications Magazine},
  vol.~53, no.~3, pp. 165--171, March 2015.

\bibitem{TS36842}
\BIBentryALTinterwordspacing
3GPP. {Study on Small Cell enhancements for E-UTRA and E-UTRAN, 2015}.
  [Online]. Available:
  \url{http://www.3gpp.org/ftp/Specs/archive/36_series/36.842/36842-c00.zip}
\BIBentrySTDinterwordspacing

\bibitem{36300}
\BIBentryALTinterwordspacing
{LTE-WLAN Aggregation and RAN Controlled LTE-WLAN Interworking, 2016}.
  [Online]. Available: \url{http://www.3gpp.org/DynaReport/36300.htm}
\BIBentrySTDinterwordspacing

\bibitem{lagrange2014very}
X.~Lagrange, ``{Very tight coupling between LTE and Wi-Fi for advanced
  offloading procedures},'' in \emph{Wireless Communications and Networking
  Conference Workshops (WCNCW)}, 2014.

\bibitem{RP150180}
\BIBentryALTinterwordspacing
Qualcomm. {Motivation for LTE-WiFi Aggregation, March 2015}. [Online].
  Available: \url{http://www.3gpp.org/DynaReport/TDocExMtg--RP-67--31196.htm}
\BIBentrySTDinterwordspacing

\bibitem{LWIR}
S.~Prashant, B.~Ajay, S.~Thomas Valerrian~Pasca, T.~Bheemarjuna~Reddy, and
  A.~Antony~Franklin, ``Lwir: Lte-wlan integration at rlc layer with virtual
  wlan scheduler for efficient aggregation,'' in \emph{Proceedings of
  GLOBECOM}.\hskip 1em plus 0.5em minus 0.4em\relax IEEE, 2016.

\bibitem{LWIP_DEMO}
S.~Thomas Valerrian~Pasca, P.~Sumanta, T.~Bheemarjuna~Reddy, and
  A.~Antony~Franklin, ``Tightly coupled lte wi-fi radio access networks: A demo
  of lwip,'' in \emph{Proceedings of COMSNETS Demo}.\hskip 1em plus 0.5em minus
  0.4em\relax IEEE, 2017.

\bibitem{lwip3gpp}
\BIBentryALTinterwordspacing
{LTE/WLAN Radio Level Integration Using IPsec Tunnel (LWIP) encapsulation;
  Protocol specification}. [Online]. Available:
  \url{http://www.3gpp.org/DynaReport/36361.htm}
\BIBentrySTDinterwordspacing

\bibitem{SWIP50}
\BIBentryALTinterwordspacing
T.~Bheemarjuna~Reddy, A.~Antony~Franklin, and S.~Thomas Valerrian~Pasca.
  {Traffic steering strategies for LTE-Wi-Fi aggregation}. [Online]. Available:
  \url{http://tsdsi.org/standards/swip/50/}
\BIBentrySTDinterwordspacing

\bibitem{git}
\BIBentryALTinterwordspacing
Thomas. {Class Diagram}. [Online]. Available:
  \url{https://github.com/ThomasValerrianPasca/C-LWIP/}
\BIBentrySTDinterwordspacing

\bibitem{ns3}
\BIBentryALTinterwordspacing
{NS-3 Simulator}. [Online]. Available: \url{https://www.nsnam.org/}
\BIBentrySTDinterwordspacing

\end{thebibliography}

\end{document}